# Numerical Simulations of Negative-index Refraction in Wedge-Shaped Metamaterials

Z. G. Dong, S. N. Zhu,* and H. Liu

*National Laboratory of Solid State Microstructures, Nanjing University, Nanjing 210093, China*

J. Zhu and W. Cao

*Material Research Institute, The Pennsylvania State University, Pennsylvania 16802, USA*

A wedge-shaped structure made of split-ring resonators (SRR) and wires is numerically simulated to evaluate its refraction behavior. Four frequency bands, namely, the stop band, left-handed band, ultralow-index band, and positive-index band, are distinguished according to the refracted field distributions. Negative phase velocity inside the wedge is demonstrated in the left-handed band and the Snell's law is conformed in terms of its refraction behaviors in different frequency bands. Our results confirmed that negative index of refraction indeed exists in such a composite metamaterial and also provided a convincing support to the results of previous Snell's law experiments.

# Ⅰ. INTRODUCTION

Left-handed materials (LHM) with simultaneously negative electric permittivity $\varepsilon$ and magnetic permeability $\mu$ have attracted a great deal of attention in recent years due to their unique electromagnetic responses which can not be expected from naturally occurring materials [1-4]. Since the first LHM consisting of metallic split ring resonators (SRR) and wires was reported [4], numerous investigations have been conducted both theoretically as well as experimentally to better understand such composite structures. Several methods have been used to verify whether or not such a proposed structure possesses negative index of refraction. The effective $\varepsilon$ and $\mu$ can be deduced by averaging the electromagnetic field in a unit cell with appropriate boundary conditions [5,6], while Smith *et al*. numerically calculated them by analyzing the scattering parameters (i.e., the transmission and reflection coefficients) [7]. In addition, there are also several alternative ways to confirm whether or not a proposed structure is an LHM, such as the Snell's law experiment [4], flat-slab imaging [8], negative beam shift [9], negative Goos-Hanchen shift [10], reversal Cherenkov radiation, and reversal Doppler shift. The most interesting property of an LHM is the negative index of refraction, i.e., the incident wave will be bent to the same side of the surface normal as that of the incident wave, while for normal materials, the beam should be refracted to the opposite side of the surface normal. Therefore, one of the most intuitive verifications of an LHM is the Snell's law experiment on wedge-shaped structures. Experimental results showed that the electromagnetic wave was indeed refracted to the same side of the refraction surface normal as that of the incident wave [4,11,12].

Previous Snell's law experiments were mostly performed in a two-dimensional scattering chamber with two parallel conducting plates to force the polarization of the electric field to be perpendicular to them and putting microwave absorbers laterally to reduce spurious reflections. However, such a measurement is not easy to perform, hence, its result has been questioned [13] because the metallic metamaterial is very lossy and there is no well-defined beam in the realized left-handed band (very narrow bandwidths in microwave region) [14]. Furthermore, although the results of these experiments [4,11,12] are affirmative, the LHM interpretation for the composite metamaterial comprising SRR and wires has been confronted by some researchers [15]. In view of the above, numerical evaluation of such Snell's law experiments is valuable since it can describe the refracted field distributions more clearly and can obtain the phase information inside the composite metamaterial. To the best of our knowledge, previously simulated wedges were mostly regarded as continuous media with given negative real parts of $\varepsilon$ and $\mu$ rather than simulating the real composite metamaterial itself [12,16,17,18]. There are other simulations on negative refraction of wedge-shaped photonic crystals in the literature [19], which are physically different from LHM principle, although both of them lead to negative refraction phenomenon. In this work, a wedge-shaped structure comprising copper SRR and wires has been directly investigated using full-wave simulation to gain a better understanding on its negative refraction behavior. Our numerical results provide as a strong complementary support to previous experimental results on negative refraction of such metamaterials from the Snell's law experiments.

## II. NUMERICAL MODEL

The unit cell of the composite metamaterial is shown in Fig. 1(a) with lattice parameters similar to those used in Refs. [12,20]. However, here we used a wider continuous wire instead of a second wire in the unit cell to generate sufficient negative $\varepsilon$ [21]. Figure 1(b) is the configuration used to simulate the negative refraction behavior of the wedge-shaped composite metamaterial. The wedge is positioned between two magnetically conducting plates (perfect magnetic boundaries). We use two magnetically conducting plates instead of two conventional electrically conducting plates because the latter requires more computer resource to reach sufficient accuracy and hence is not computationally efficient. The spacing between these two plates is 3.3 mm (one unit cell). It is not necessary to stack unit cells along the *x*-direction in the simulation provided the periodic nature is represented by appropriate periodic boundary conditions. For the wedge model, 12 unit cells (39.6 mm) are used along the *y*-axis and 7 unit cells (23.1 mm) are used along the *z*-axis. The refraction interface has a staircase pattern with two-unit-cell step in the *y*-direction and one-unit-cell step in the *z*-direction, which can be referred to as a wedge of 26.6° since the wedge can be approximately treated as an effectively homogeneous medium for wave propagation when the wavelengths are larger than the unit cell size. The semicircular vacuum chamber with a radius of 4.2 cm is the refraction section (refraction section with larger radius does not influence the refraction direction, but reduces the calculation accuracy under the limitations of computer resource and run time, as shown in Fig. 5). The rectangular vacuum channel is used to guide the electromagnetic wave to be incident normally onto the first interface of the wedge and then propagates into the wedge. Such a one-dimensional composite metamaterial will exhibit LHM behavior only if the incident wave is polarized in an appropriate direction with respect to the SRR and wires. Therefore, the incident wave is guided along the rectangular channel with magnetic field in the $x$-direction (perpendicular to the SRR plane) and electric field in the $y$-direction (along the wire).

## III. RESULTS AND DISCUSSION

### A. Frequency bands

Figure 2 is the magnitude distribution of the electric field at the top magnetically conducting plate. Four frequency regions are distinguished within the frequency range from 13 GHz to 25 GHz. The First frequency region from 13 GHz to 16 GHz is the stop band in which $\varepsilon_y$ is negative while $\mu_x$ is positive. Weak intensity transmission has been found in this stop band [Figure 2(a) at 14 GHz], which is caused by the surface wave running along the wedge [See Fig. 4(a), note that both $\varepsilon_z$ and $\mu_x$ of the wedge are positive in this band]. The second frequency region from 16 GHz to 18 GHz is the left-handed band, where the polarized incident microwave is refracted negatively due to the double negative nature of $\varepsilon_y$ and $\mu_x$ in this band. Figure 2(b) is a typical field distribution for this band at 16.6 GHz. The third frequency region from 18 GHz to 20 GHz is the ultralow-index band, in which the effective refractive index is ultralow so that the refraction direction is nearly parallel to the normal of the refraction interface [Figure 2(c) at 20 GHz]. Generally, this band is referred to as the stop band with negative $\varepsilon_y$ and positive $\mu_x$. However, in our simulations, this band occurs right at the falling edge of the permittivity versus frequency curve (according to the S-parameters simulation, the plasma frequency of this structure is at about 20 GHz). Therefore, the amplitude of $\varepsilon_y$ in this band is too small (although negative) to completely block the transmission of the incident wave [21]. Nevertheless, the transmission magnitude in this band is not as strong as in its neighboring frequency regions (see Fig. 3). The fourth frequency region from 20 GHz to 25 GHz is the positive-index band where both $\varepsilon_y$ and $\mu_x$ are positive. The effective refractive index in this band increases with frequency from ultralow to near unity [Figure. 2(d) at 25 GHz]. This is consistent with the theoretical results of Refs. [7,22]. Based on our simulation results, one can make two additional remarks: (1) the reflections at the two interfaces of the wedge cannot be neglected because the wedge is not impedance matched to the vacuum chamber; (2) the metallic metamaterial is very lossy, especially for waves with frequencies inside in the left-handed band.

Figure 3 is the surface plots of the magnitude of the electric field around a semicircle of 4 cm in radius (the concentric semicircular edge of the refraction chamber has a radius of 4.2 cm), from which the overall field distribution as a function of refraction angle and frequency can be visualized. The transmitted intensity is strongly attenuated in the left-handed band as compared with that of the positive-index band. The larger loss is attributed to the resonant nature of the composite metamaterial in the left-handed band.

**B. Phase information inside the composite metamaterial**

It is well known that the propagation of an electromagnetic wave in an LHM is characterized by its negative phase velocity; in other words, the electromagnetic wave appears to move backward. Therefore, if the SRR-and-wire metamaterial indeed has LHM behavior as it has been claimed [4], negative phase velocity must be observed when microwaves propagate through the wedge. In our simulations, negative phase velocity inside the wedge is demonstrated in any plane parallel to the $z$-$y$ plane when the propagating wave frequencies are in the left-handed band. Figure 4 is the field distribution at the bottom magnetically conducting plate at three frequencies. Backward phase propagation is indeed observed at 16.6GHz as shown by the arrows representing the phase propagation direction (plots in planes adjacent to the metallic elements gives less obvious phase information due to serious resonant concentration of the field around the metallic elements). Our simulation results provided a solid support to the LHM behavior of the composite metamaterial consisting of SRR and wires.

**C. Snell's law simulation**

The Snell's law is the basic description of the refraction phenomenon at a boundary between two media with different indices of refraction. The Snell's is also applicable to materials with negative refraction indices. In this work, three wedges with different angles have been used to confirm that the Snell's law is obeyed in the metamaterials being investigated. As shown in Fig. 5, the refraction chamber in this simulation has been extended to the dimensions of 15 cm × 15 cm in order to obtain clearer pictures of the refraction angles. Three wedges with wedge angles of 18.4°, 26.6°, and 45° have been simulated, which correspond to refraction angles of -7°, -10°, and -15°, respectively, for a 16.6 GHz incident wave. In Fig. 6, $\sin\theta_t$ as a function of $\sin\theta_i$ is plotted at three different frequencies (16.6 GHz, 20 GHz, and 25 GHz) and a linear dependence of these two quantities are clearly demonstrated as in the Snell's law, where $\theta_i$ and $\theta_t$ are the angles of incident and refracted waves with respect to the normal of the incident interface. One can see from Fig. 6 that the incident wave is refracted in a manner consistent with the Snell's law. Consequently, by taking the average of the three wedges, the effective refractive indices of the simulated wedge can be calculated to be -0.38, 0.005, and 0.90, corresponding to the wave frequencies of 16.6 GHz, 20 GHz, and 25 GHz, respectively.

## IV. SUMMARY AND CONCLUSION

In summary, we report here our simulation results on a wedge-shaped composite metamaterial. Negative phase velocity inside the metamaterial is indeed exhibited inside the left-handed band in which the effective $\varepsilon$ and $\mu$ are both negative. Furthermore, simulations on three wedges of different angles have confirmed that the Snell's law is obeyed in the metamaterials having negative refraction indices. Our simulation results clearly revealed the LHM nature of the metamaterials and provided a strong support to those Snell's law experiments conducted previously.

## ACKNOWLEDGMENTS

This work was supported by the State Key Program for Basic Research of China (Grant No. 2004CB619003), and by the National Natural Science Foundation of China under contract No. 90201008.

FIG. 1. (Color online) (a) Unit cell of the composite metamaterial used in the simulation. C = 0.25 mm, D = 0.3 mm, G = 0.46 mm, W = 2.63 mm, E = 0.5 mm, and L = 3.3 mm. The metallic elements are copper with a thickness of 0.017 mm. The SRR is square and the unit cell is cubic. (b) The simulated configuration with a wedge angle of 26.6°. The electromagnetic wave propagates along the *z*-axis, with electric field along the *y*-axis and magnetic field along the *x*-axis.

FIG. 2. (Color online) The magnitude distribution of the electric field in the top magnetically conducting plate at four frequencies: (a) 14 GHz; (b) 16.6 GHz; (c) 20 GHz; (d) 25 GHz.

FIG. 3. (Color online) Surface plot of the electric field magnitude around a 4 cm radius semicircle.

FIG. 4. (Color online) The magnitude distribution of the electric field in the bottom magnetically conducting plate at three different frequencies to demonstrate the phase velocity, which is indicated by arrows: (a) 14 GHz. The antiparallel arrow in the incident section represents strong reflection. (b) 16.6 GHz, with negative phase velocity in the wedge. (c) 25 GHz, with positive phase velocity in the wedge.

FIG. 5. (Color online) Refraction field distributions at 16.6 GHz for three different wedge angles: (a) 18.4°, (b) 26.6°, and (c) 45°. Larger refraction chamber with 15 cm × 15 cm dimensions is used to obtain clearer pictures of the refraction directions.

FIG. 6. Snell's law verification by three wedges at different frequencies. The angles that the incident and refracted waves make to the normal of the interface are distinguished by $\theta_i$ and $\theta_t$, respectively.

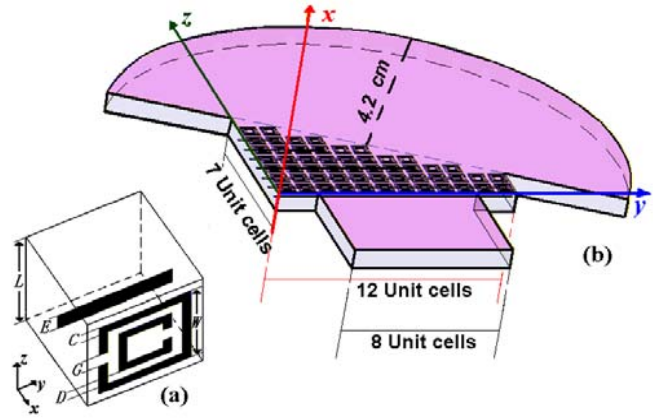

Figure 1. Z. G. Dong *et al.*

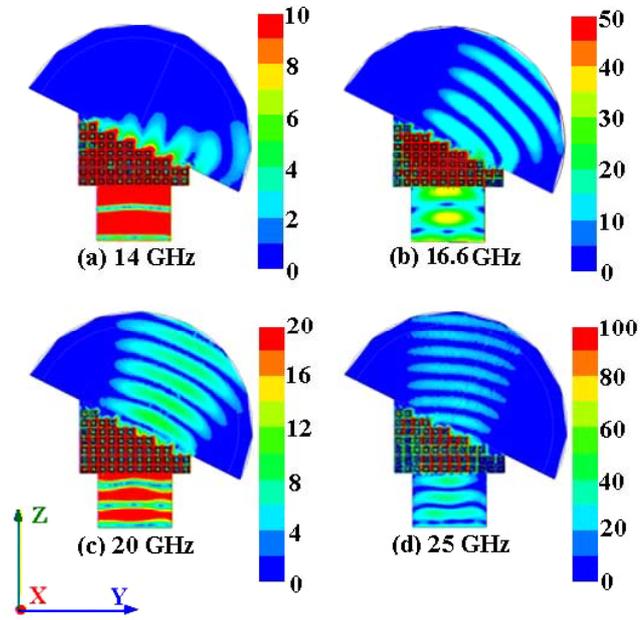

Figure 2. Z. G. Dong *et al.*

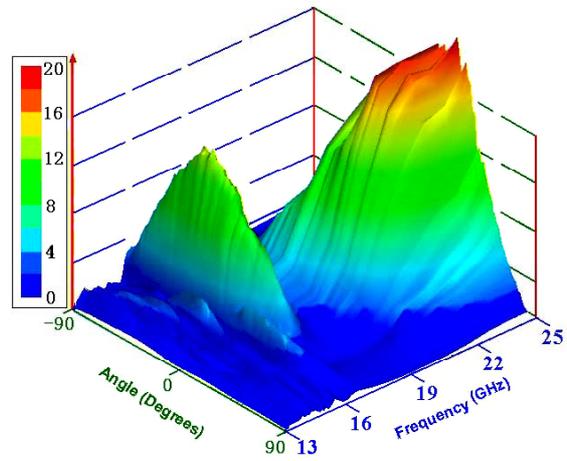

Figure 3. Z. G. Dong *et al.*

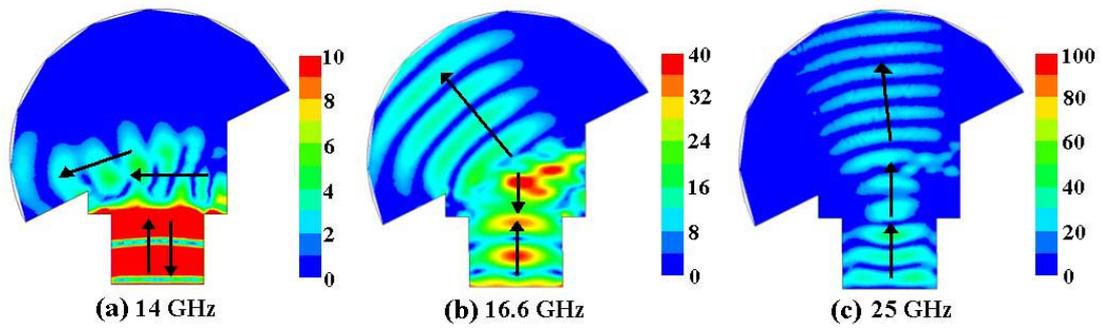

Figure 4. Z. G. Dong *et al.*

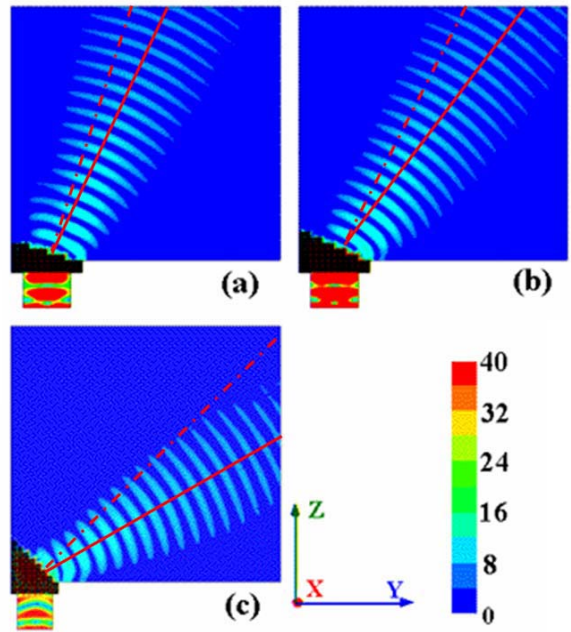

Figure 5. Z. G. Dong *et al.*

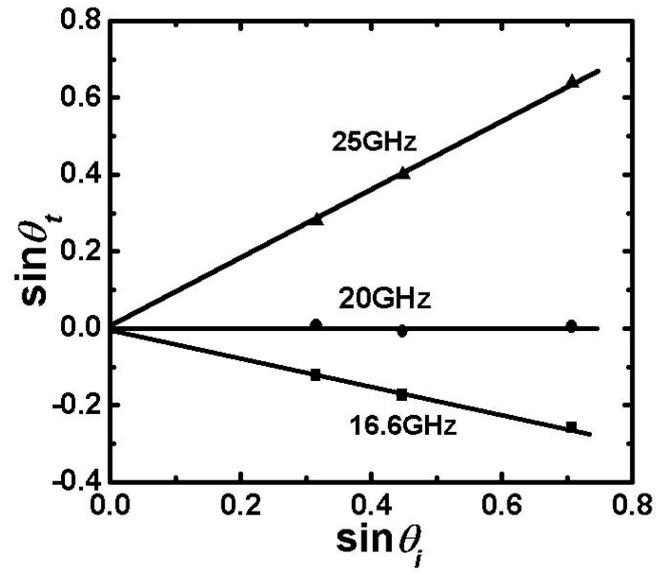

Figure 6. Z. G. Dong *et al.*